\documentclass[11pt,titlepage,a4paper]{article}
\usepackage{bozhomac}	
\usepackage{bozhlogo}

\usepackage{xr1}
\title{\bfseries	\vspace*{-2.1095in}
\vspace*{-3ex}
{
\begin{flushright}
	  \textbf{\large LANL xxx E-print archive No. quant-ph/9806046}\\[2ex]
\end{flushright}
}
{\huge Fibre bundle formulation of	\\[6pt]
nonrelativistic quantum mechanics	\\[2ex]
\Large III. Pictures and integrals of motion
}
}

\vspace{3ex}

\author{
Bozhidar Z. Iliev
\thanks{Department Mathematical Modeling,
Institute for Nuclear Research and \mbox{Nuclear} Energy,
Bulgarian Academy of Sciences,
Boul. Tzarigradsko chauss\'ee~72, 1784 Sofia, Bulgaria}
\thanks{E-mail address: bozho@inrne.bas.bg}
\thanks{URL: http://www.inrne.bas.bg/mathmod/bozhome/}
}

\date{
\vspace{2.27ex}\ShortTitle{Bundle quantum mechanics: III}	\\[0.27ex]
\vspace{3.27ex}
	\begin{tabular}{r@{$\colon\to~$}l}
\vspace{0.09ex} Basic ideas	& March 1996		\\[0.09ex]
\vspace{0.09ex} Began		& May 19, 1996		\\[0.09ex]
\vspace{0.09ex} Ended		& July 12, 1996		\\[0.09ex]
\vspace{0.09ex} Revised		& December 1996 -- January 1997,\\[0.09ex]
\vspace{0.09ex} Revised		& April 1997, September 1998	\\[0.09ex]
\vspace{0.09ex} Last update	& October 29, 1998  	\\[0.09ex]
\vspace{0.09ex} Composing/Extracting part III
			& September 27/October 4, 1997	\\[0.09ex]
\vspace{0.09ex} Updating part III   & October 29, 1998	\\[0.09ex]
\vspace{0.27ex} Produced	& \fbox{\today}		\\[0.27ex]
	\end{tabular} \\[1.27ex]
	\begin{tabular}{r@{$\colon~$}l}
\vspace{0.27ex} LANL xxx archive server E-print No.& quant-ph/9806046
							\\[0.27ex]
	\end{tabular} \\[-0.27ex]
\vspace{3.27ex}{\Huge\BOZHO}	\\[3.27ex]
\vspace{0.27ex}\Subject{Quantum mechanics; Differential geometry} \\[2.27ex]
	\begin{tabular}{r@{\hspace{0.512em}}|@{\hspace{0.512em}}l}
\vspace{0.27ex}		  
&
\vspace{0.27ex}
	\end{tabular} \\[1.27ex]
\vspace{0.27ex}	\\[0.27ex]
}
\newcommand{\Hil}{\mathcal{F}}	
\newcommand{\HilB}{(\bHil,\proj,\base)}	
	\newcommand{\bHil}{\mathit{F}}	
	\newcommand{\proj}{\pi}		
	\newcommand{\base}{\mathit{M}}	

\newcommand{\Ham}{\mathcal{H}}	
\newcommand{\bHam}{\mathit{H}}	

\newcommand{\mHam}{\boldsymbol{\Ham}^\mathbf{m}}   
\newcommand{\mbHam}{\boldsymbol{\bHam}\!^\mathbf{m}}

\newcommand{\dyn}[1]{\pmb{\mathbb{#1}}}	
	\newcommand{\ope}[1]{\mathcal{#1}}		 
	\newcommand{\mor}[1]{\mathit{#1}}		 
	\newcommand{\mmor}[1]{\boldsymbol{\mathit{#1}}}	 

\newcommand{\ih}{\mathrm{i}\hbar}
\newcommand{\iih}{\frac{1}{\ih}} 

\begin{document}		


\renewcommand*{\thesection}{I.\arabic{section}}
\renewcommand*{\thefootnote}{\protect{I.\arabic{footnote}}}
\setcounter{page}{0}
\setcounter{footnote}{0}
\setcounter{section}{0}
\setcounter{equation}{0}
 	\include{bqm-1txt}
\renewcommand*{\thesection}{\arabic{section}}
\renewcommand*{\thefootnote}{\arabic{footnote}}
\setcounter{page}{0}
\setcounter{footnote}{0}
\setcounter{section}{0}
\setcounter{equation}{0}

\renewcommand*{\thesection}{II.\arabic{section}}
\renewcommand*{\thefootnote}{\protect{II.\arabic{footnote}}}
\setcounter{page}{0}
\setcounter{footnote}{0}
\setcounter{section}{0}
\setcounter{equation}{0}
 	\include{bqm-2txt}
\renewcommand*{\thesection}{\arabic{section}}
\renewcommand*{\thefootnote}{\arabic{footnote}}
\setcounter{page}{0}
\setcounter{footnote}{0}
\setcounter{section}{0}
\setcounter{equation}{0}


\renewcommand{\thefootnote}{\fnsymbol{footnote}}
\maketitle			
\renewcommand{\thefootnote}{\arabic{footnote}}

\tableofcontents		


\pagestyle{myheadings}
\markright{\itshape\bfseries Bozhidar Z. Iliev:
	\upshape\sffamily\bfseries Bundle quantum mechanics.~III}

\begin{abstract}

We propose a new systematic fibre bundle formulation of nonrelativistic
quantum mechanics. The new form of the theory is equivalent to the usual one
but it is in harmony with the modern trends in theoretical physics and
potentially admits new generalizations in different directions. In it
a pure state of some quantum system is described by a state section (along
paths) of a (Hilbert) fibre bundle. It's evolution is determined through the
bundle (analogue of the) Schr\"odinger equation. Now the dynamical variables
and the density operator are described via bundle morphisms (along paths).
The mentioned quantities are connected by a number of relations derived in
this work.
%
%

	In this third part of our series we investigate the bundle analogues
of the conventional pictures of motion. In particular, there are found the
state sections and bundle morphisms corresponding to state vectors and
observables respectively. The equations of motion for these quantities are
derived too. Using the results obtained, we consider from the bundle
view-point problems concerning the integrals of motion. An invariant (bundle)
necessary and sufficient conditions for a dynamical variable to be an integral
of motion are found.

\end{abstract}

\section {Introduction}
\label{introduction-III}
\setcounter{equation} {0}

	The present paper is a third part of our series on fibre bundle
formulation of nonrelativistic quantum mechanics. It is a direct continuation
of~\cite{bp-BQM-introduction+transport,bp-BQM-equations+observables}.

	The work is organized in the following way.

	The bundle description of the different pictures of motion is
presented in Sect.~\ref{VII}. The Schr\"odinger picture, which, in fact,
was investigated
in~\cite{bp-BQM-introduction+transport,bp-BQM-equations+observables},
is reviewed in
Subsection~\ref{VII.1}. To the bundle Heisenberg picture is devoted
Subsection~\ref{VII.2}. The corresponding equations of motion for the
observables are derived and discussed. In Subsection~\ref{VII.3} is
investigated the `general' picture of motion obtained by means of an
arbitrary linear unitary bundle map (along paths). In it are derived
and discussed different equations for the state sections and
observables.

	In Sect.~\ref{VIII} are investigated problems concerning the
integrals of motion from fibre bundle point of view. An interesting result
here is that a dynamical variable is an integral of motion iff the
corresponding to it bundle morphism is transported along the observer's world
line with the help of the (bundle) evolution transport.

	Sect~\ref{conclusion-III} closes the paper.
\vspace{1.2ex}

	The notation of the present work is the the same as the one
in~\cite{bp-BQM-introduction+transport,bp-BQM-equations+observables}
and will not be repeated here.

	The references to sections, equations, footnotes etc.
from~\cite{bp-BQM-introduction+transport,bp-BQM-equations+observables} are
obtained from their corresponding sequential reference numbers
in~\cite{bp-BQM-introduction+transport,bp-BQM-equations+observables} by
adding in
front of them the Roman one (I) or two (II), respectively, and a dot as a
separator.  For instance, Sect.~I.5 and (II.2.7) mean respectively section 5
of~\cite{bp-BQM-introduction+transport} and equation~(2.7) (equation 7 in
Sect.~2) of~\cite{bp-BQM-equations+observables}.

	Below, for reference purposes, we present a list of some essential
equations of~\cite{bp-BQM-introduction+transport,bp-BQM-equations+observables}
which are used in this paper.  Following the above convention, we retain
their original reference numbers.
\enlargethispage*{6ex}
	\begin{gather}
\tag{\ref{2.1}}
\psi(t_2) = \ope{U}(t_2,t_1)\psi(t_1),
\\
\tag{\ref{2.5}}
\ih\frac{d \psi(t)}{d t} = \Ham(t) \psi(t),
\\
\tag{\ref{2.7}}
\Ham(t) =
\ih\frac{\partial \ope{U}(t,t_0)}{\partial t} \circ \ope{U}^{-1}(t,t_0) =
\ih\frac{\partial \ope{U}(t,t_0)}{\partial t} \circ \ope{U}(t_0,t),
\\
\tag{\ref{2.9}}
\langle\ope{A}(t)\rangle_\psi^t :=
\frac{\langle\psi(t) | \ope{A}(t)\psi(t)\rangle}
{\langle\psi(t) | \psi(t)\rangle},
\\
\tag{\ref{4.3a}}
\Psi_\gamma(t) = l_{\gamma(t)}^{-1}(\psi(t))\in \bHil_{\gamma(t)},
\\
\tag{\ref{4.11}}
\langle \mor{A}_{x\to y}^{\ddag} \Phi_x| \Psi_y \rangle_y :=
\langle \Phi_x| \mor{A}_{y\to x}\Psi_y \rangle_x,
\qquad \Phi_x\in \bHil_x,\quad \Psi_y\in \bHil_y,
\\
\tag{\ref{4.2b}}
\ope{U}^\dag(t_1,t_2) = \ope{U}^{-1}(t_2,t_1),
\\
\tag{\ref{4.4}}
\Psi_\gamma(t) = \mor{U}_\gamma(t,s) \Psi_\gamma(s),
\\
\tag{\ref{4.7}}
\mor{U}_\gamma(t,s) =
		l_{\gamma(t)}^{-1}\circ \ope{U}(t,s) \circ l_{\gamma(s)},
\qquad s,t\in J,
\\
\tag{\ref{4.14}}
\mor{U}_{\gamma}^{\ddag}(t,s) = \mor{U}_\gamma(t,s) =
				\mor{U}_{\gamma}^{-1}(s,t),
	\end{gather}
	\begin{gather}
\tag{\ref{5.1}}
\ih\frac{d \boldsymbol{\Psi}_\gamma(t) }{dt} =
\mbHam_{\gamma}(t) \boldsymbol{\Psi}_\gamma(t),
\\
\tag{\ref{5.10}}
\boldsymbol{\Gamma}_\gamma(t) :=
\left[ \Gamma_{{\ }a}^{b}(t;\gamma) \right] =
- \iih \mbHam_{\gamma}(t),
\\
\tag{\ref{5.13}}
\mor{D}_{t}^{\gamma}\Psi_\gamma = 0,
\\
\tag{\ref{6.0}}
\mor{A}_{\gamma}(t) =
l_{\gamma(t)}^{-1} \circ \ope{A}(t) \circ l_{\gamma(t)}
\colon  \bHil_{\gamma(t)}\to\bHil_{\gamma(t)},
\\
\tag{\ref{6.1'}}
\left\langle
\mor{A}_{\gamma}(t)
\right\rangle_{\Psi_\gamma}^{t} =
\frac{
\langle \Psi_\gamma(t) | \mor{A}_\gamma(t) \Psi_\gamma(t) \rangle_{\gamma(t)}
}
{
\langle \Psi_\gamma(t) | \Psi_\gamma(t) \rangle_{\gamma(t)}
},
\\
\tag{\ref{6.1''}}
\langle \ope{A}(t)\rangle_{\psi}^{t} =
\langle \mor{A}_{\gamma}(t) \rangle_{\Psi_\gamma}^{t},
\\
\tag{\ref{6.5}}
\pmb{\boldsymbol{[}} \tilde \mor{D}_{t}^{\gamma} (C) \pmb{\boldsymbol{]}} =
\frac{d}{dt} \boldsymbol{C}_t +
\left[
\boldsymbol{\Gamma}_\gamma(t),\boldsymbol{C}_t
\right]_{\_} .
	\end{gather}

\section {Pictures of motion from bundle view-point}
\label{VII}
\setcounter{equation} {0}

	Well-known are the different pictures (or representations) of
motion of a quantum system~\cite[ch.~VIII,~\S\S~9, 10, 14]{Messiah-QM},
\cite[ch.~III,~\S~14]{Fock-FQM}, \cite[\S~27, \S~28]{Dirac-PQM}:
the Schr\"odinger's, Heisenberg's, interaction's, and other
`intermediate' ones. Bellow we consider certain problems connected with
these special representations of the motion of a quantum system from the
developed here fibre bundle view-point on quantum mechanics.

\subsection {Schr\"odinger picture}
\label{VII.1}
\setcounter{equation} {0}

	In fact the (bundle) Schr\"odinger picture of motion of a
quantum system is the way of its description we have been dealing
until now~\cite{bp-BQM-introduction+transport,bp-BQM-equations+observables}.

	Let \(\gamma\colon J\to \base\) be the world line of an observer. The
state of a system is described by a (state) section
\(
\Psi_\gamma\in \Sec\left(\left.\HilB\right|_{\gamma(J)}\right)
\)
which, possibly, may be multiple-valued;
at the moment $t$ it is represented by the state vector
\(
\Psi_\gamma(t) := \left.\Psi_\gamma\right|_{\gamma(t)}
\).
It is generally a variable in time quantity evolving according
to the bundle Schr\"odinger equation~(\ref{5.13}).

	In this description to any dynamical variable $\dyn{A}$ there
corresponds a unique bundle morphism $\mor{A}$ along paths over $\HilB$.
With respect to the observer at $\gamma(t)$ it reduces to $\mor{A}_\gamma(t)$
which  generally depends on $t$ nevertheless that in the Hilbert space
description the observable can be time-independent.  It (or its evolution) is
explicitly given by~(\ref{6.0}).

	Geometrically the bundle morphisms of $\HilB$ `live' in the
\emph{fibre bundle $\morf\HilB:=(\bHil_0,\proj_0,\base)$ of bundle morphisms}
of $\HilB$ introduce
by~\cite[equations~(3.9) and~(3.10)]{bp-TP-morphisms}:
	\begin{equation}	\label{7.1}
	\begin{array}{l}
\bHil_0 =
  \{ \mor{B}_x:\quad \mor{B}_x\colon \bHil_x\to \bHil_x,\ x\in \base \}, \\
\proj_0(\mor{B}) = x_\mor{B},\quad \mor{B}\in \bHil_0,\
	\end{array}
	\end{equation}
for the unique $x_\mor{B}\in \base$ for which
$\mor{B}\colon \bHil_{x_\mor{B}}\to \bHil_{x_\mor{B}}$.

	Evidently
\(
\mor{B}\in  \Morf\HilB \iff
\mor{B}\in  \Sec\left(\morf\HilB\right),
\)
i.e.\
	\begin{equation}	\label{7.2}
\Morf\HilB = \Sec\left(\morf\HilB\right).
	\end{equation}

	A morphism $\mor{A}$ along paths is represented with respect to
a given observer by, maybe multiple-valued, morphism
\(
\mor{A}_\gamma\in\Morf\left(\left.\HilB\right|_{\gamma(J)}\right)
	 = \Sec\left(\morf\HilB|_{\gamma(J)}\right)
\)
such that
\(
\left.\mor{A}_\gamma\right|_{\gamma(t)} :=
	 \{ \mor{A}_\gamma(s) : s\in J, \quad \gamma(s)=\gamma(t) \}.
\)

	Summing up, in the bundle Schr\"odinger picture (in the general
case) both the state vectors and the dynamical variables change with
time in the above-described way in the corresponding fibre bundles.

\subsection {Heisenberg picture}
\label{VII.2}

	Below we present two different ways for introduction of bundle
Heisenberg picture leading, of course, to one and the same result. The
first one is based  entirely on the bundle approach and reveals its natural
geometric character. The second one is a direct analogue of the usual way in
which one arrives to this picture.

	According to~\cite[sect.~\ref{+:Sect4}]{bp-normalF-LTP}
or~\cite[sect.~3]{bp-LTP-general}
\emph{%
any linear transport along paths is locally Euclidean},
i.e.\ (see~\cite[sect.~\ref{+:Sect4}]{bp-normalF-LTP} for details and
rigorous results)
along any path there is a field of
(generally multiple-valued~\cite[remark~4.2]{bp-normalF-LTP})
bases, called normal, in which its matrix is unit.
In particular, along $\gamma\colon  J\to \base$ there exists a set of bases
\(
\left\{
\{ \tilde e_{a}^{\gamma}(t)\}\mathrm{\ -\ basis\ in\ } \bHil_{\gamma(t)}
\right\}
\)
in which the matrix of the bundle evolution transport $\mor{U}_\gamma(t,s)$ is
\(
   \widetilde{\mmor{U}}_\gamma(t,s)=\openone.
\)
	Explicitly one can put
\(
   \tilde e_{a}^{\gamma}(t) = \mor{U}_\gamma(t,t_0) e_{a}^{\gamma}(t_0),
\)
where $t,t_0\in J$ and the basis $\{ e_{a}^{\gamma}(t_0) \}$ in
$\bHil_{\gamma(t_0)}$ is
fixed~\cite[proof of proposition~3.1]{bp-LTP-general}
(cf.~\cite[equation~\eref{+:4.2}]{bp-normalF-LTP}).%
\footnote{%
The so-defined bases are not uniquely defined at the points of
self-intersection, if any, of $\gamma$. Evidently, they are unique on
any `part' of $\gamma$ without self-intersections. The last case
covers the interpretation of $\gamma$ as an observer's world line, in
which it cannot have self-intersections.
See~\cite[sect.~4]{bp-normalF-LTP} for details.%
}
Because of~(\ref{4.6}), (\ref{5.9}), and~(\ref{5.10}) this class of
special bases along $\gamma$ is uniquely defined by any one of the
(equivalent) equalities:
	\begin{equation}	\label{7.3}
\widetilde{\mmor{U}_\gamma}(t,t_0) = \openone,	\quad
\widetilde{\boldsymbol{\Gamma}}_\gamma(t) = \boldsymbol{0},	\quad
\widetilde{\mbHam_\gamma}(t) = \boldsymbol{0}.
	\end{equation}

	So, in such a special basis the matrix-bundle Hamiltonian
vanishes and, consequently (see~(\ref{5.1})), the components of the
state vectors remain constant in time $t$, i.e.\
\( \widetilde{\boldsymbol{\Psi}}_\gamma(t)=\const, \)
but the vectors themselves are not necessary such.

	In
\( \{  \tilde e_{a}^{\gamma}(t) \} \)
the components of $\mor{A}_\gamma(t)$ are
	\begin{align*}
\widetilde{(\mor{A}_\gamma(t))}_{ab} &=
\left\langle
\tilde e_{a}^{\gamma}(t) |
\left( \left.\mor{A}_\gamma\right|_{\bHil_{\gamma(t)}}\right)
			\tilde e_{b}^{\gamma}(t)
\right\rangle_{\gamma(t)}
\\ &=
\langle
\mor{U}_\gamma(t,t_0) e_{a}^{\gamma}(t_0) |
\mor{A}_\gamma(t) \mor{U}_\gamma(t,t_0) e_{b}^{\gamma}(t_0)
\rangle_{\gamma(t)}
\\ &=
\langle
e_{a}^{\gamma}(t_0) |
\mor{U}_\gamma^{-1}(t,t_0)\mor{A}_\gamma(t)
		\mor{U}_\gamma(t,t_0) e_{b}^{\gamma}(t_0)
\rangle_{\gamma(t)} =
\widetilde{\left( \mor{A}_{\gamma,t}^{\bHam}(t_0) \right)}_{ab},
	\end{align*}
where
	\begin{equation}	\label{7.4}
\mor{A}_{\gamma,t}^{\mathrm{H}}(t_0) :=
\mor{U}_\gamma^{-1}(t,t_0)\circ \mor{A}_\gamma(t)\circ \mor{U}_\gamma(t,t_0)
\colon  \bHil_{\gamma(t_0)}\to  \bHil_{\gamma(t_0)}.
	\end{equation}
Hence the matrix elements of $\mor{A}_\gamma(t)$ in
$\{\tilde e_{a}^{\gamma}(t)\}$
coincide with those of $\mor{A}_{\gamma,t}^{\mathrm{H}}(t_0)$ in
$\{e_{a}^{\gamma}(t_0)\}.$
	Consequently, due to~(\ref{2.9}), (\ref{6.1'}), (\ref{6.1''}),
and~(\ref{4.14}),
the mean value of $\mor{A}$ (along $\gamma$) is
	\begin{equation*}\begin{split}
\langle \mor{A}_\gamma(t)\rangle_{\Psi_\gamma}^{t} &=
\left(\widetilde{\mor{A}_\gamma(t)}\right)_{ab}
\widetilde{\Psi}_{\gamma}^{a}(t) \widetilde{\Psi}_{\gamma}^{b}(t) /
\langle
{\Psi}_{\gamma}(t) | {\Psi}_{\gamma}(t)
\rangle_{\gamma(t)}
\\ &=
\Bigl(  \widetilde{\mor{A}_{\gamma,t}^{\mathrm{H}}(t_0)}  \Bigr)_{ab}
\widetilde{\Psi}_{\gamma}^{a}(t) \widetilde{\Psi}_{\gamma}^{b}(t) /
\langle
{\Psi}_{\gamma}(t_0) | {\Psi}_{\gamma}(t_0)
\rangle_{\gamma(t)} .
	\end{split}\end{equation*}
But
\(
\widetilde{\boldsymbol{\Psi}}_\gamma(t) =
\widetilde{\boldsymbol{\Psi}}_\gamma(t_0),
\)
hereout
	\begin{equation}	\label{7.5}
\langle \mor{A}_\gamma(t) \rangle_{\Psi_\gamma}^{t} =
\langle \mor{A}_{\gamma,t}^{\mathrm{H}}(t_0) \rangle_{\Psi_\gamma}^{t_0}.
	\end{equation}

	So, the mean value of $\mor{A}_\gamma(t)$ in a state $\Psi_\gamma(t)$
is equal to the mean value of $\mor{A}_{\gamma,t}^{\mathrm{H}}(t_0)$ in the
state $\Psi_\gamma(t_0)$. Taking into account that the only measurable
(observable) physical quantities are the mean
values~\cite{Dirac-PQM,Messiah-QM,Neumann-MFQM}, we infer that the
descriptions of a quantum system through either one of the pairs
$( \Psi_\gamma(t),\mor{A}_\gamma(t) )$
and
$( \Psi_\gamma(t_0),\mor{A}_{\gamma,t}^{\mathrm{H}}(t_0) )$
are fully equivalent. The former one is the Schr\"odinger picture of
motion, considered above in Sect.~\ref{VII.1}. The latter one is the
\emph{ bundle Heisenberg picture of motion}
 of the quantum system. In it the time dependence of the
state vectors is entirely shifted to the observables in conformity
with~(\ref{7.4}). In this description the (bundle) state vectors are constant
and do not evolve in time. On the contrary, in it the observables depend on
time and act on one and the same fibre of $\HilB$, the one to which belongs
the (initial) state vector. Their evolution is governed by the Heisenberg
form of the bundle Schr\"odinger equation~(\ref{5.13}) which can be derived
in the following way.

	Substituting~(\ref{4.7}) and~(\ref{6.0}) into~(\ref{7.4}),
we get
	\begin{equation}	\label{7.6}
\mor{A}_{\gamma,t}^{\mathrm{H}}(t_0) =
l_{\gamma(t_0)}^{-1}\circ\ope{A}_{t}^{\mathrm{H}}(t_0)\circ l_{\gamma(t_0)}
\colon \bHil_{\gamma(t_0)}\to \bHil_{\gamma(t_0)},
	\end{equation}
where (cf.~(\ref{7.4}))
	\begin{equation}	\label{7.7}
\ope{A}_{t}^{\mathrm{H}}(t_0) :=
\ope{U}(t_0,t)\circ\ope{A}(t)\circ\ope{U}(t,t_0)\colon \Hil\to\Hil
	\end{equation}
is the Heisenberg operator corresponding to $\ope{A}(t)$ in the Hilbert space
description (see below).

	A simple verification shows that
	\begin{equation}	\label{7.8}
\ih\frac{\partial \ope{A}_t^{\mathrm{H}}(t_0)}{\partial t} =
\left[  \ope{A}_t^{\mathrm{H}}(t_0),\Ham_t^{\mathrm{H}}(t_0)  \right] _{\_} +
\ih\left(\frac{\partial\ope{A}}{\partial t}\right)_{\!t}^{\!\mathrm{H}} (t_0).
	\end{equation}
Here
\( \left({\partial\ope{A}}/{\partial t}\right)_t^{\mathrm{H}} (t_0) \)
is obtained from~(\ref{7.7}) with ${\partial\ope{A}}/{\partial t}$ instead of
$\ope{A}$ and
	\begin{equation}	\label{7.9}
\Ham_t^{\mathrm{H}}(t_0) =
\ope{U}^{-1}(t,t_0) \Ham(t) \ope{U}(t,t_0) =
\ih\ope{U}^{-1}(t,t_0)\frac{\partial\ope{U}(t,t_0)}{\partial t}
	\end{equation}
(cf.~(\ref{7.7})) with $\Ham(t)$ being the usual Hamiltonian in $\Hil$
(see~(\ref{2.7})), i.e.\ $\Ham_t^{\mathrm{H}}(t_0)$ is the Hamiltonian in the
Heisenberg picture.

	Finally, from~(\ref{7.6}) and~(\ref{7.8}), we obtain
	\begin{equation}	\label{7.10}
\ih\frac{\partial \mor{A}_{\gamma,t}^{\mathrm{H}}(t_0)}{\partial t} =
\left[
\mor{A}_{\gamma,t}^{\mathrm{H}}(t_0), H_{\gamma,t}^{\mathrm{H}}(t_0)
\right] _{\_} +
\ih\left(
\frac{\partial\ope{A}}{\partial t}
\right)_{\!\gamma,t}^{\!\mathrm{H}} (t_0)
	\end{equation}
in which all quantities with subscript $\gamma$ are defined according
to~(\ref{7.6}). This is the bundle equation of motion (for the observables)
in the Heisenberg picture of motion of a quantum system. It determines the
time evolution of the observables in this description.

	Now we shall outline briefly how the above results can be obtained by
transferring the (usual) Heisenberg picture of motion from the Hilbert
space $\Hil$ to its analogue in the fibre bundle $\HilB$.

	The mathematical expectation of an observable $\ope{A}(t)$ in a
state with a
state vector $\psi(t)$ is (see~(\ref{2.9}), (\ref{2.4}), and~(\ref{4.2b}))
\[
\langle\ope{A}(t)\rangle_\psi^t 	=
\frac{\langle\psi(t) | \ope{A}(t)\psi(t)\rangle}
{\langle\psi(t) | \psi(t)\rangle} =
\frac{\langle\psi(t_0) |
	\ope{U}^{-1}(t,t_0)\ope{A}(t)\ope{U}(t,t_0)\psi(t_0)\rangle}
{\langle\psi(t_0) | \psi(t_0)\rangle}	.
\]
	Combining this with~(\ref{7.7}), we find:
	\begin{gather}	\label{7.11}
\langle\ope{A}(t)\rangle_\psi^t 	=
\langle\ope{A}_t^{\mathrm{H}}(t_0)\rangle_{\psi}^{t_0} 	=
\langle\ope{A}_t^{\mathrm{H}}(t_0)\rangle_{\psi_t^{\mathrm{H}}}^{t_0},
\\ \label{7.11a}
\psi_t^{\mathrm{H}}(t_0) := \psi(t_0).
	\end{gather}

	Thus the pair $(\psi(t),\ope{A}(t))$ is equivalent to the pair
$(\psi(t_0),\ope{A}_t^{\mathrm{H}}(t_0))$ from the view-point of observable
quantities. The latter one realizes the Heisenberg picture in $\Hil$. In it
the state vectors are constant while the observables, generally,
change with time according to the Heisenberg form~(\ref{7.8}) of the
equation of motion.

	The bundle morphisms along paths corresponding to $\ope{A}$ and
$\ope{A}_t^{\mathrm{H}}(t_0)$ are defined (see~(\ref{6.0})), respectively, by
\(
\mor{A}_\gamma(t)=l_{\gamma(t)}^{-1}\circ\ope{A}(t)\circ l_{\gamma(t)}
\)
and (see~(\ref{7.7}) and~(\ref{4.7}))
	\begin{equation}	\label{7.12}
\mor{A}_{\gamma,t}^{\mathrm{H}}(t_0)	=
l_{\gamma(t_0)}^{-1}\circ\ope{A}_t^{\mathrm{H}}(t_0)\circ l_{\gamma(t_0)}	=
\mor{U}_\gamma^{-1}(t,t_0)\circ \mor{A}_\gamma(t)\circ \mor{U}_\gamma(t,t_0).
	\end{equation}
Hence to the Heisenberg operator $\ope{A}_t^{\mathrm{H}}$
corresponds exactly the introduced
above by~(\ref{7.4}) (Heisenberg) morphism
$\mor{A}_{\gamma,t}^{\mathrm{H}}(t_0)$.
In particular, to the Hamiltonian $\Ham(t)$ and its Heisenberg form
$\Ham_t^{\mathrm{H}}(t_0)$, given by~(\ref{7.7}) for $\ope{A}=\Ham$ or
by~(\ref{7.9})
(cf.~(\ref{2.7})), correspond the morphisms
(see~(\ref{6.0}) and~(\ref{6.2'}))
\(
\bHam_\gamma(t)	=
l_{\gamma(t)}^{-1}\circ\Ham(t)\circ l_{\gamma(t)}
\)
and (cf.~(\ref{7.6}) and~(\ref{7.9}))
	\begin{equation}	\label{7.13}
\bHam_{\gamma,t}^{\mathrm{H}}(t_0)	=
l_{\gamma(t_0)}^{-1}\circ\Ham_t^{\mathrm{H}}(t_0)\circ l_{\gamma(t_0)}	=
\mor{U}_\gamma^{-1}(t,t_0)\circ H_\gamma(t)\circ \mor{U}_\gamma(t,t_0)
	\end{equation}
the latter of which is exactly the one entering in~(\ref{7.10}).

	Now it is a trivial verification that the morphisms
$\mor{A}_{\gamma,t}^{\mathrm{H}}(t_0)$ satisfy the bundle Heisenberg
equation of motion~(\ref{7.10}).

	Thus the both approaches are self-consistent and lead to
one and the same final result, the bundle Heisenberg picture of motion.

	According to the above discussion, in the bundle Heisenberg
picture the state of a quantum system is described by a constant
state vector
	\begin{equation}	\label{7.13a}
\Psi_{\gamma,t}^{\mathrm{H}}(t_0) =
			\Psi_{\gamma}(t_0) \in \bHil_{\gamma(t_0)}
	\end{equation}
and generally  changing in time observables
\( \mor{A}_{\gamma,t}^{\mathrm{H}}(t_0)\in\proj_0^{-1}(\gamma(t_0)) \)
(see the notation at the end of subsection~\ref{VII.1}).
If $\partial \ope{A}/\partial t =0$, which is assumed
usually~\cite{Dirac-PQM,Messiah-QM},
our results depend only on the point $x=\gamma(t)$ but not on the map
$\gamma$ itself; otherwise, as for different $\gamma$ the point
$x=\gamma(t)$  may be different,
the theory depends explicitly on the observer's world line
$\gamma\colon J\to \base$ and it (or, more precisely, the bundle $\HilB$)
has to be restricted on the set $\gamma(J)$.

	In the Hilbert bundle description the transition from the
Schr\"odinger picture to the Heisenberg one is by means of~\eqref{7.4}
and~\eqref{7.13a}, while in the Hilbert space description it is
via~\eqref{7.7} and~\eqref{7.11a}. A feature of the former description is
that in it the Heisenberg picture is `locally' identical with the
Schr\"odinger one in a special field of bases characterized by any one of the
equations in~\eqref{7.3}.

	An interesting interpretation of the Heisenberg picture can be given
in the fibre bundle $\morf{\HilB}=(\bHil_0,\proj_0,\base)$ of bundle
morphisms of $\HilB$ (see subsection~\ref{VII.1}). Since $\mor{U}$ is a
transport along paths in the fibre bundle $\HilB$, then, according
to~\cite[equation~(3.12)]{bp-TP-morphisms}, it induces a transport
$^\circ{\!}\mor{U}$ along paths in $\morf{\HilB}$ whose action on a morphism
$\mor{A}_\gamma$ along $\gamma\colon J\to\base$ is
	\begin{equation}	\label{7.13b1}
^\circ{\!}\mor{U}_\gamma(t,s) (\mor{A}_\gamma(s)) :=
\mor{U}_\gamma(t,s)\circ \mor{A}_\gamma(s)\circ \mor{U}_\gamma(s,t)
\in\proj_{0}^{-1}(\gamma(t)).
	\end{equation}

	Comparing this definition with~\eqref{7.4}, we obtain
	\begin{equation}	\label{7.13b2}
^\circ{\!}\mor{U}_\gamma(t_0,t) (\mor{A}_\gamma(t)) :=
\mor{A}_{\gamma,t}^{\mathrm{H}}(t_0).
	\end{equation}
Consequently the transport $^\circ{\!}\mor{U}$ along paths is just the map
which maps the bundle Schr\"odinger representation of the observables into
their bundle Heisenberg representation. A simple corollary of this fact
and~\eqref{3.1} is that the Heisenberg morphisms along paths, like
\(
\mor{A}_{t}^{\mathrm{H}}\colon\gamma\mapsto\mor{A}_{\gamma,t}^{\mathrm{H}}
\colon x\mapsto
\{ \mor{A}_{\gamma,t}^{\mathrm{H}}(s) : s\in J,\ \gamma(s)=x \}
\),
are linearly (or, more precisely, $^\circ{\!}\mor{U}$-)transported sections
along paths of $\morf{\HilB}$:
	\begin{equation}	\label{7.13b3}
\mor{A}_{\gamma,t}^{\mathrm{H}}(t_1)  =
^\circ{\!}\mor{U}_{\gamma,t}(t_1,t_0) \mor{A}_{\gamma,t}^{\mathrm{H}}(t_0).
	\end{equation}

	Let $^\circ{\!}\mor{D}$ be the derivation along paths corresponding to
$^\circ{\!}\mor{U}$ according to~\eqref{5.7} (see
also~\cite{bp-normalF-LTP,bp-LTP-general}), i.e.\
\(
   ^\circ{\!}\mor{D}\colon\gamma\mapsto{^\circ{\!}\mor{D}^\gamma}
   \colon s\mapsto {^\circ{\!}\mor{D}_s^\gamma}
\)
with
	\begin{equation}	\label{7.13b4}
^\circ{\!}\mor{D}_s^\gamma (\mor{A}_\gamma) :=
\lim_{\varepsilon\to0}
\left\{
\frac{1}{\varepsilon}
\left[
^\circ{\!}\mor{U}(s,s+\varepsilon)
\left( \mor{A}_\gamma(s+\varepsilon) \right) - \mor{A}_\gamma(s)
\right]
\right\}  .
	\end{equation}

	A simple calculation shows that in a local field of bases the matrix
of $^\circ{\!}\mor{D}_s^\gamma (\mor{A}_\gamma)$ is
	\begin{equation}	\label{7.13b5}
\pmb{\boldsymbol{[}}
^\circ{\!}\mor{D}_s^\gamma (\mor{A}_\gamma)
\pmb{\boldsymbol{]}} =
-\left[
\mmor{A}_\gamma(s),\boldsymbol{\Gamma}_\gamma(s)
\right]_{\_}
+
\frac{\partial \mmor{A}_\gamma(s)}{\partial s}
	\end{equation}
where
\(
\boldsymbol{\Gamma}_\gamma(s) :=
\left[\Gamma_{\ a}^{b}(s;\gamma)\right]:=
\left.\partial\mmor{U}_\gamma(s,t) /\partial t \right|_{t=s}
\)
is the matrix of the coefficients of $\mor{U}$
(not of $^\circ{\!}\mor{U}$!).
From here, using~\eqref{5.10} and~\eqref{6.2''}, one, after some matrix
algebra, finds the explicit form of~\eqref{7.13b4}:
	\begin{equation}	\label{7.13b6}
^\circ{\!}\mor{D}_t^\gamma (\mor{A}_\gamma) =
\iih [ \mor{A}_\gamma(t) , \mor{H}_\gamma(t) ]_{\_} +
\left(\frac{\partial \ope{A}}{\partial t}\right)_{\!\gamma(t)},
	\end{equation}
the last term being defined via~\eqref{6.0}.

The last result, together with~\eqref{7.4}, shows that the Heisenberg
equation of motion~\eqref{7.10} is equivalent to
	\begin{equation}	\label{7.13b7}
\frac{\partial \mor{A}_{\gamma,t}^{\mathrm{H}}(t_0) } {\partial t} =
\mor{U}_\gamma(t_0,t)\circ
\left(
^\circ{\!}\mor{D}_t^\gamma ( \mor{A}_{\gamma} )
\right)
\circ \mor{U}_\gamma(t,t_0).
	\end{equation}
By the way, this equation is also an almost trivial corollary
of~\eqref{7.13b4}, \eqref{7.13b2}, and~\eqref{3.1}.

	Now the analogue of~\eqref{5.14} is
	\begin{equation}	\label{7.13b8}
^\circ{\!}\mor{D}_t^\gamma \circ ^\circ{\!}\mor{U}^\gamma(t,t_0) = 0,
\quad
^\circ{\!}\mor{U}^\gamma(t_0,t_0) = \id_{\proj_{0}^{-1}(\gamma(t_0))}.
	\end{equation}
From here and~\eqref{7.13b3}, we derive the
\emph{bundle Heisenberg equation of motion}
(for the observables) as
	\begin{equation}	\label{7.13b9}
^\circ{\!}\mor{D}_{t_0}^\gamma
\left( \mor{A}_{\gamma,t}^{\mathrm{H}}(t_0) \right) =0
	\end{equation}
which is another equivalent form of~\eqref{7.10}.

\subsection {`General' picture}
\label{VII.3}

	Let the map
\(
\mor{V}_\gamma(t,s)\colon \bHil_{\gamma(s)}\to \bHil_{\gamma(t)},\ s,t\in J
\)
be linear and unitary, i.e.\ (see~(\ref{4.12d}))
$\mor{V}_{\gamma}^{\ddag}(t,s)=\mor{V}_{\gamma}^{-1}(s,t)$,
where $\mor{V}_{\gamma}^{-1}(s,t)$ is the \emph{left} inverse of
$\mor{V}_{\gamma}(s,t)$.
A simple calculation shows that
	\begin{equation}	\label{7.14}
\langle \mor{A}_\gamma(t) \rangle_{\Psi_\gamma}^{t} =
\langle \mor{A}_{\gamma,t}^\mor{V}(t_1)
				\rangle_{\Psi_{\gamma,t}^\mor{V}}^{t_1} ,
	\end{equation}
where~(\ref{4.11}) was used, $t_1\in J$, and
	\begin{align}	\label{7.15}
\Psi_{\gamma,t}^\mor{V}(t_1) &:=
\mor{V}_\gamma(t_1,t)\Psi_\gamma(t)\in \bHil_{\gamma(t_1)},
\\				\label{7.16}
\mor{A}_{\gamma,t}^\mor{V}(t_1) &:=
\mor{V}_\gamma(t_1,t) \circ \mor{A}_\gamma(t) \circ
\mor{V}_\gamma^{-1}(t_1,t)\colon
\bHil_{\gamma(t_1)} \to \bHil_{\gamma(t_1)}.
	\end{align}

	Thereof the pairs $(\Psi_\gamma(t),\mor{A}_\gamma(t))$
and $(\Psi_{\gamma,t}^\mor{V}(t_1),\mor{A}_{\gamma,t}^\mor{V}(t_1))$
provide a fully equivalent description of a given quantum system.
The latter description  can be called the
\emph{$\mor{V}$-picture} or \emph{general picture}

of motion. For $t_1=t$ and $\mor{V}_\gamma(t,t)=\id_{\bHil_{\gamma(t)}}$ it
coincides with the Schr\"odinger picture and for $t_1=t_0$ and
$\mor{V}_\gamma(t_0,t)=\mor{U}_\gamma(t_0,t)$
it reproduces the Heisenberg picture.

	The equations of motion in the $\mor{V}$-picture cannot be obtained
directly by differentiating~(\ref{7.15}) and~(\ref{7.16}) with
respect to $t$ because derivatives like
$\partial \mor{V}_\gamma(t_1,t)/\partial t$
are not (`well') defined due to
$\mor{V}_\gamma(t_1,t)\colon \bHil_{\gamma(t)} \to  \bHil_{\gamma(t_1)} $.
They can be derived by differentiating the corresponding
to~(\ref{7.15}) and~(\ref{7.16}) matrix equations, but we prefer the
below-described method which explicitly reveals the connections between
the conventional and the bundle descriptions of  quantum evolution.

	The analogues of~(\ref{7.14}), (\ref{7.15}), and~(\ref{7.16}) in the
Hilbert space $\Hil$, which is the typical fibre of $\HilB$, are
respectively:
	\begin{align}	\label{7.17}
\langle \ope{A}(t) \rangle_{\psi}^{t} &=
\langle \ope{A}_{t}^{\mor{V}}(t_1) \rangle_{ \psi_{t}^{\mor{V}} }^{t_1}
\left( = \langle \mor{A}_\gamma(t) \rangle_{\Psi_\gamma}^{t} \right),
\\	\label{7.18}
\psi_{t}^{\ope{V}}(t_1) &:= \ope{V}(t_1,t)\psi(t) \in \Hil,
\\	\label{7.19}
\ope{A}_{t}^{\ope{V}}(t_1) &:=
\ope{V}(t_1,t) \circ \ope{A}(t) \circ \ope{V}^{-1}(t_1,t) \colon  \Hil\to\Hil
	\end{align}
where $\ope{V}(t_1,t) \colon  \Hil\to\Hil$ is the linear unitary
(i.e.\ $\ope{V}^\dag(t_1,t) = (\ope{V}(t,t_1))^{-1}$) operator corresponding
to
\( \mor{V}(t_1,t)\colon \bHil_{\gamma(t)}\to \bHil_{\gamma(t_1)} \),
that is we have (cf.~(\ref{4.7}))
	\begin{equation}	\label{7.20}
\mor{V}_\gamma(t_1,t)
= l_{\gamma(t_1)}^{-1}\circ \ope{V}(t_1,t)\circ l_{\gamma(t)}.
	\end{equation}

	 The description of the quantum evolution via
$\psi_{t}^{\ope{V}}(t_1)$ and $\ope{A}_{t}^{\ope{V}}(t_1)$
is the $\mor{V}$-picture of motion in $\Hil$. Besides,
due to~(\ref{4.3a}), (\ref{6.0}), and~(\ref{7.15})--(\ref{7.20}), we have:
	\begin{align}	\label{7.21}
\Psi_{\gamma,t}^{\ope{V}}(t_1) &:=
  l_{\gamma(t_1)}^{-1} \left( \psi_{t}^{\ope{V}}(t_1) \right) =
\Psi_{\gamma,t}^{\mor{V}}(t_1),
\\	\label{7.22}
\mor{A}_{\gamma,t}^{\ope{V}}(t_1)   &:=
l_{\gamma(t_1)}^{-1} \circ \ope{A}_{t}^{\ope{V}}(t_1) \circ l_{\gamma(t_1)} =
\mor{A}_{\gamma,t}^{\mor{V}}(t_1).
	\end{align}
According to~(\ref{7.20})--(\ref{7.22}) the sets of
equalities~(\ref{7.14})--(\ref{7.16}) and~(\ref{7.17})--(\ref{7.19})
are equivalent; they are, respectively, the Hilbert bundle and the
(usual) Hilbert space descriptions of the $\mor{V}$-picture of motion.

	Differentiating~(\ref{7.18}) with respect to $t$, substituting
into the so-obtained result the Schr\"odinger equation~(\ref{2.5}),
and introducing the modified Hamiltonian
	\begin{align}	\label{7.23}
\widetilde{\Ham}(t) &:= \Ham(t) - \,_\ope{V}\!\Ham(t_1,t),
\\	\label{7.23a}
_\ope{V}\! \Ham(t_1,t) &:=
\ih \frac{\partial\ope{V}^{-1}(t_1,t)}{\partial t} \circ \ope{V}(t_1,t) =
-\ih \ope{V}^{-1}(t_1,t) \circ \frac{\partial\ope{V}(t_1,t)}{\partial t},
	\end{align}
we find the state vector's equation of motion in the $\mor{V}$-picture as
	\begin{equation}	\label{7.24}
\ih \frac{\partial\psi_{t}^{\ope{V}}(t_1)}{\partial t} =
\widetilde{\Ham}_{t}^{\ope{V}}(t_1) \psi_{t}^{\ope{V}}(t_1),
	\end{equation}
where
	\begin{equation}	\label{7.25}
\widetilde{\Ham}_{t}^{\ope{V}}(t_1) =
\ope{V}(t_1,t) \circ \widetilde{\Ham}(t) \circ \ope{V}^{-1}(t_1,t) =
\Ham_{t}^{\ope{V}}(t_1) - \,_\ope{V}\!\Ham_{t}^{\ope{V}}(t_1)
	\end{equation}
with
	\begin{equation}	\label{7.26}
	\begin{split}
\Ham_{t}^{\ope{V}}(t_1) &:=
	\ope{V}(t_1,t) \circ {\Ham}(t) \circ \ope{V}^{-1}(t_1,t),
\\
\,_\ope{V}\!\Ham_{t}^{\ope{V}}(t_1) &:=
\ope{V}(t_1,t) \circ _\ope{V}\!\Ham(t_1,t) \circ \ope{V}^{-1}(t_1,t) =
-\ih \frac{\partial \ope{V}(t_1,t)}{\partial t} \circ \ope{V}^{-1}(t_1,t)
	\end{split}
	\end{equation}
is the $\mor{V}$-form of~(\ref{7.23}).

	The \emph{equation of motion for the observables in the
$\mor{V}$-picture} in $\Hil$ is obtained in an analogous way.
Differentiating~(\ref{7.19})  with respect to $t$ and applying~(\ref{7.26}),
we find

	\begin{equation}	\label{7.27}
\ih \frac{\partial\ope{A}_{t}^{\ope{V}}(t_1)}{\partial t} =
\left[
\ope{A}_{t}^{\ope{V}}(t_1), _\ope{V}\!\Ham_{t}^{\ope{V}}(t_1)
\right]_{\_}
+
\ih\left( \frac{\partial\ope{A}(t)}{\partial t} \right)
_{\!t}^{\!\ope{V}} (t_1) .
	\end{equation}

	Now the bundle equations of motion in the $\mor{V}$-picture are
simply a corollary of the already obtained ones in $\Hil$. In fact,
differentiating the first equalities from~(\ref{7.21}) and~(\ref{7.22})
with respect to $t$ and then using~(\ref{7.24}), (\ref{7.27}),
(\ref{7.21}), and~(\ref{7.22}), we, respectively, get:
	\begin{align}	\label{7.28}
\ih \frac{\partial \Psi_{\gamma,t}^{\mor{V}}(t_1) }{\partial t}  &=
\widetilde{\bHam}_{\gamma,t}^{\mor{V}}(t_1) \Psi_{\gamma,t}^{\mor{V}} (t_1),
\\	\label{7.29}
\ih \frac{\partial \mor{A}_{\gamma,t}^{\mor{V}}(t_1) }{\partial t}     &=
\left[
\mor{A}_{\gamma,t}^{\mor{V}}(t_1), _\mor{V}\!\bHam_{\gamma,t}^{\mor{V}}(t_1)
\right]_{\_}
+
\ih
\left(
\frac{\partial \ope{A}(t)} {\partial t}
\right)_{\!\gamma,t}^{\!\mor{V}}(t_1),\ \ \ \ \
	\end{align}
where
	\begin{equation}	\label{7.28-29}
	\begin{split}
\widetilde{\bHam}_{\gamma,t}^{\mor{V}}(t_1) &=
l_{\gamma(t_1)}^{-1} \circ \widetilde{\Ham}_{t}^{\ope{V}}(t_1) \circ
l_{\gamma(t_1)} =
\mor{V}_\gamma(t_1,t) \circ \widetilde{\bHam}_\gamma(t) \circ
\mor{V}_{\gamma}^{-1}(t_1,t) ,
\\
_\mor{V}\!\bHam_{\gamma,t}^{\mor{V}}(t_1) &=
l_{\gamma(t_1)}^{-1} \circ  _\ope{V}\!\Ham_{t}^{\ope{V}}(t_1) \circ
l_{\gamma(t_1)} =
\mor{V}_\gamma(t_1,t) \circ _\ope{V}\!\bHam_\gamma(t_1,t) \circ
\mor{V}_{\gamma}^{-1}(t_1,t),
	\end{split}
	\end{equation}
with
\(
\widetilde{\bHam}_\gamma(t) :=
l_{\gamma(t)}^{-1} \circ \widetilde{\Ham}(t)  \circ l_{\gamma(t)}
\)
and
\(
_\ope{V}\!\bHam_\gamma(t_1,t) =
-\ih
\mor{V}^{-1}(t_1,t) \circ l_{\gamma(t_1)} \circ
\frac{\partial \ope{V}(t_1,t)}{\partial t} \circ l_{\gamma(t)},
\)
are the modified and the `additional' Hamiltonians in the $\mor{V}$-picture
(cf.~(\ref{7.23}), (\ref{7.16}), and~(\ref{7.22})).

	Now we consider briefly the evolution operator and transport in the
$\mor{V}$-picture. In $\Hil$ and in $\HilB$ they are define respectively
by (cf.~(\ref{2.1}) and~(\ref{4.4}))
	\begin{align}	\label{7.30}
\psi_{t}^{\ope{V}}(t_1) &=
		\ope{U}^\ope{V}(t,t_1,t_0)\psi_{t_0}^{\ope{V}}(t_1),
\\	\label{7.31}
\Psi_{\gamma,t}^{\mor{V}}(t_1 ) &=
\mor{U}_\gamma^\mor{V}(t,t_1,t_0)\Psi_{\gamma,t_0}^{\mor{V}}(t_1),
	\end{align}
and, due to~(\ref{7.24}) and~(\ref{7.28}), satisfy the
following initial-value problems:
	\begin{align}	\label{7.32}
\ih \frac{\partial \ope{U}^\ope{V}(t,t_1,t_0) }{\partial t} &=
\widetilde{\Ham}_{t}^{\ope{V}}(t_1)\circ\ope{U}^\ope{V}(t,t_1,t_0),&
\ope{U}^\ope{V}(t_0,t_1,t_0) &= \id_\Hil,
\\				\label{7.33}
\ih \frac{\partial \mor{U}_\gamma^\mor{V}(t,t_1,t_0) }{\partial t} &=
\widetilde{\bHam}_{\gamma,t}^{\mor{V}}(t_1)\circ
				\mor{U}_\gamma^\mor{V}(t,t_1,t_0),
&
\mor{U}_\gamma^\mor{V}(t_0,t_1,t_0) &= \id_{\bHil_{\gamma(t_1)}}.
	\end{align}

	Combining~(\ref{7.30}) , (\ref{7.18}), and~(\ref{2.1})
from one hand and~(\ref{7.31}), (\ref{7.15}), and~(\ref{4.4}) from another
hand, we respectively obtain:
	\begin{align}	\label{7.34}
\ope{U}^\ope{V}(t,t_1,t_0) &= \ope{V}(t_1,t)\circ \ope{U}(t,t_0)\circ
				\ope{V}^{-1}(t_1,t_0)\colon \Hil\to\Hil,
\\			\label{7.35}
\mor{U}_\gamma^\mor{V}(t,t_1,t_0) &=
\mor{V}_\gamma(t_1,t)\circ \mor{U}_\gamma(t,t_0)\circ
				\mor{V}_\gamma^{-1}(t_1,t_0)
\colon \bHil_{\gamma(t_1)}\to \bHil_{\gamma(t_1)}.
	\end{align}

	Notice that in the Heisenberg picture we have
	\begin{equation}	\label{7.35a}
\ope{U}^\mathrm{H}(t,t_0,t_0) =  \id_\Hil \mbox{ and }
\mor{U}_\gamma^\mathrm{H}(t,t_0,t_0) =  \id_{\bHil_{\gamma(t_0)}},
	\end{equation}
respectively.

	Substituting in~(\ref{7.35}) the equalities~(\ref{7.20})
and~(\ref{4.7}) and taking into account~(\ref{7.34}), we find the
connection between the two evolution operators in the $\mor{V}$-picture as
	\begin{equation}	\label{7.36}
\mor{U}_\gamma^\mor{V}(t,t_1,t_0) =
l_{\gamma(t_1)}^{-1}\circ \ope{U}^\ope{V}(t,t_1,t_0)  \circ l_{\gamma(t_1)}.
	\end{equation}

	At the end of this subsection we want to notice that the
derived here equations have a direct practical applications in
connection with the approximate treatment of the problem of quantum
evolution of state vectors and observables
(cf.~\cite[ch.~VIII, \S~14]{Messiah-QM}). Indeed,
by~(\ref{7.23}) we have
$\Ham(t) = _\ope{V}\!\Ham(t,t_1) + \widetilde{\Ham}(t) $.
We can consider
	\begin{equation}	\label{7.37}
\Ham^{(0)}(t) := _\ope{V}\!\Ham(t_1,t) =
\ih \frac{\partial \ope{V}^{-1}(t_1,t)}{\partial t} \circ \ope{V}(t_1,t)
	\end{equation}
as a given approximate (unperturbed) Hamiltonian of the quantum system with
evolution operator
$\ope{U}^{(0)}(t_1,t) = \ope{V}(t_1,t)$.
(In this case $\bHam^{(0)}(t)$ is independent of $t_1$ and
$\ope{V}^{-1}(t_1,t) = \ope{V}(t,t_1).$)
	Then $\widetilde{\Ham}(t)$ may be regarded, in some `good'
cases, as a `small' correction to $\bHam^{(0)}(t)$. In other words, we can
say that $\bHam^{(0)}(t)$ is the Hamiltonian of the `free' system, while
$\Ham(t)$ is its Hamiltonian when a given interaction with Hamiltonian
$\widetilde{\Ham}(t)$ is introduced.

	In this interpretation the $\mor{V}$-picture is the well known
\emph{interaction picture}.
In it one supposes to be given the basic (zeroth order) Hamiltonian
$\Ham^{(0)}(t) := \mbox{$_\ope{V}\!\Ham(t,t_1)$}$
and the interaction Hamiltonian
$\Ham^{(I)}(t) = \widetilde{\Ham}(t)$.
On their base can be computed all other quantities of the system described by
them. In particular, all of the above results hold true for
\(
\ope{V}(t_1,t) = \ope{U}^{(0)}(t_1,t) =
\Texp\left(
\int\limits_{t}^{t_1} \Ham^{(0)}(\tau) d\tau / \ih \right)
\).
	Besides, in this case the total evolution operator
\(
\ope{U}(t,t_0) =
\Texp\left(
\int\limits_{t_0}^{t}\Ham(\tau) d\tau / \ih \right)
\)
splits into
	\begin{equation}	\label{7.38}
\ope{U}(t,t_0) =
\ope{U}^{(0)}(t,t_0) \circ \ope{U}^{(I)}(t,t_0)
	\end{equation}
with
\(
\ope{U}^{(I)}(t,t_0) :=
\Texp\left(
\int\limits_{t_0}^{t}
\left(\Ham^{(I)}\right)_{\tau}^{\ope{U}^{(0)}(t_0)}
d\tau / \ih \right),
\)
where
$\left(\Ham^{(I)}\right)_{\tau}^{\ope{U}^{(0)}(t_0)}$
is an operator given by~(\ref{7.25}) for
$\widetilde{\Ham}=\Ham^{(I)}$, $t_1=t_0$,
and $\ope{V}(t_0,t)=\ope{U}^{(0)}(t_0,t)$.
Now the equations of motion~(\ref{7.24}) and~(\ref{7.27}) take,
respectively, the form:
	\begin{align}	\label{7.39}
\ih \frac{\partial \psi^{(I)}(t)}{\partial t} &=
\left(\Ham^{(I)}\right)_{ \!\ope{U}^{(0)} }^{\! t}(t_0) \psi^{(I)}(t),
\\	\label{7.40}
\ih \frac{\partial \ope{A}^{(I)}(t)}{\partial t} &=
\left[
\ope{A}^{(I)}(t),
\left(\Ham^{(I)}\right)_{ \!\ope{U}^{(0)} }^{\! t}(t_0)
\right]_{\_}
+ \ih
\left(
\frac{\partial \ope{A}}{\partial t}
\right)^{\! \ope{U}^{(0)} }_{\! t} {\!} (t_0)
	\end{align}
where
$\psi^{(I)}(t):=\psi^{\ope{U}^{(0)} }_{t} (t_0)$
and
$\ope{A}^{(I)}(t):=\ope{A}^{\ope{U}^{(0)} }_{t} (t_0)$.
	Up to notation the last two equations coincide respectively with
equations~(55) and~(56) of~\cite[ch.VIII, \S~15]{Messiah-QM}.

	The bundle form of the interaction interpretation of the
$\mor{V}$-picture of motion will not be presented here as an almost evident
one.

\section {Integrals of motion}
\label{VIII}
\setcounter{equation} {0}

	Usually~\cite[ch.~VIII, \S~12]{Messiah-QM},
\cite[\S~28]{Dirac-PQM} an
\emph{explicitly not depending on time}
dynamical variable is called and an integral (or a constant) of motion
if the corresponding to it observable is time-independent in the
Heisenberg picture of motion. Due to~(\ref{7.8}) this means
	\begin{equation}	\label{8.1}
0 = \ih \frac{\partial \ope{A}_{t}^{\mathrm{H}}(t_0)}{\partial t} =
\left[ \ope{A}_{t}^{\mathrm{H}}(t_0) ,
			\Ham_{t}^{\mathrm{H}}(t_0) \right]_{\_} .
	\end{equation}
Hence, if
\(
{\partial \ope{A}(t)}/{\partial t} = 0,
\)
then $\ope{A}$ is an integral of motion if and only if it commutes with the
Hamiltonian. Due to~(\ref{7.7}), (\ref{7.19}), and~(\ref{7.27}) this
result holds true in any picture of motion.

	If~(\ref{8.1}) holds, then
$\partial\ope{A}(t)/\partial t = 0 $ and~(\ref{7.7}) imply
	\begin{equation}	\label{8.2}
\ope{A}(t)=\ope{A}(t_0)=\ope{A}_{t}^{\mathrm{H}}(t_0)=
					\ope{A}_{t_0}^{\mathrm{H}}(t_0).
	\end{equation}
 From~(\ref{7.7}) and~(\ref{8.2}) one easily obtains
that~(\ref{8.1}) (under the assumption
\(
{\partial \ope{A}(t)}/{\partial t} = 0
\))
is equivalent to the commutativity of the observable and the evolution
operator:
	\begin{equation}	\label{8.3}
[ \ope{A}(t_0) , \ope{U}(t_0,t) ]_{\_} = 0
	\end{equation}
which, in connection with further generalizations, is better to be
written as
	\begin{equation}	\label{8.3a}
 \ope{A}(t_0) \circ\ope{U}(t_0,t) = \ope{U}(t_0,t)\circ\ope{A}(t).
	\end{equation}

	It is almost evident that the mean values of the integrals of
motion are constant:
	\begin{equation}	\label{8.4}
\langle\ope{A}(t)\rangle_{\psi}^{t} =
\langle\ope{A}_{t}^{\mathrm{H}}(t_0)\rangle_{\psi}^{t_0} =
\langle\ope{A}_{t_0}^{\mathrm{H}}(t_0)\rangle_{\psi}^{t_0} =
\langle\ope{A}(t_0)\rangle_{\psi}^{t_0} .
	\end{equation}
In particular, if $\psi^{\mathrm{H}}(t)=\psi(t_0)$ is an eigenvector of
$\ope{A}_{t}^{\mathrm{H}}(t_0)$ with eigenvalue $a$, i.e.\
\(
\ope{A}_{t}^{\mathrm{H}}(t_0)\psi^{\mathrm{H}}(t) = a \psi^{\mathrm{H}}(t),
\)
then $a=\const$ as
$\langle \ope{A}_{t}^{\mathrm{H}}(t_0) \rangle_{\psi^{\mathrm{H}}}^{t_0} = a$.
Besides, in the Schr\"odinger picture we have
$\ope{A}(t_0)\psi(t)=a\psi(t).$

	Evidently, the identity map $\id_\Hil$, which plays the r\^ole of
the unit operator in $\Hil$, is an integral on motion. For it any state
vector is an eigenvector with $1\in\mathbb{R}$ as eigenvalue.

	Before looking on the integrals of motion from the fibre bundle
point of view, we shall generalize the above material in the general
case when
\(	{\partial \ope{A}(t)}/{\partial t}	\)
may be different from zero.

	We call a dynamical variable, which  may be explicitly
time-dependent, an
\emph{integral (or a constant) of motion}
if its mean value is time-independent. According to~(\ref{6.1''}),
(\ref{7.14}) and~(\ref{7.17}) this definition does not depend on the used
concrete picture of motion, as well as on the conventional or bundle
description of
the theory. Hence, without a lost of generality, we consider at first the
Schr\"odinger picture in $\Hil$.

	So, by definition,
$\ope{A}(t)\colon \Hil\to\Hil$ is an integral of motion if
	\begin{equation}	\label{8.5}
\langle\ope{A}(t)\rangle_{\psi}^{t} = \langle\ope{A}(t_0)\rangle_{\psi}^{t_0}
	\end{equation}
for some given instant of time $t_0$.

	Due to~(\ref{2.9}), (\ref{2.1}), (\ref{2.4}), and~(\ref{7.7}) the last
equation is equivalent to
	\begin{equation}	\label{8.6}
\ope{A}(t)=
\ope{U}(t,t_0)\circ\ope{A}(t_0)\circ\ope{U}(t_0,t) =
					\ope{A}_{t_0}^{\mathrm{H}}(t)
	\end{equation}
or to
	\begin{equation}	\label{8.7}
\ope{U}(t_0,t)\circ\ope{A}(t)=\ope{A}(t_0)\circ\ope{U}(t_0,t).
	\end{equation}
Thus~(\ref{8.3a}) remains true in the general case, when it generalizes the
commutativity of an observable and the evolution operator; in fact, in this
case we can say, by definition, that $\ope{A}$ and $\ope{U}$ commute
iff~(\ref{8.7}) holds.

	Differentiating~(\ref{8.6}) with respect to $t$ and
using~(\ref{2.7}), we see that $\ope{A}$ is an integral of motion iff
	\begin{equation}	\label{8.8}
\ih \frac{\partial \ope{A}(t)}{\partial t} +[\ope{A}(t),\Ham(t)]_{\_} = 0
	\end{equation}
which for ${\partial \ope{A}(t)}/{\partial t}=0$ reduces to~(\ref{8.1}).
In fact, according to~(\ref{7.7}) and~(\ref{7.8}), in the Heisenberg
picture~(\ref{8.8}) is equivalent to
	\begin{equation}	\label{8.9}
0 =
\ih \left( \frac{\partial \ope{A}(t)}{\partial t}
					\right)_{\! t}^{\mathrm{H}}(t_0) +
 \left[ \ope{A}_{t}^{\mathrm{H}}(t_0) ,
			\Ham_{t}^{\mathrm{H}}(t_0) \right]_{\_} =
\ih \frac{\partial \ope{A}_{t}^{\mathrm{H}}(t_0)}{\partial t}
	\end{equation}
which proves our assertion. Besides, from~(\ref{8.9}) follows
	\begin{equation}	\label{8.10}
\ope{A}_{t}^{\mathrm{H}}(t_0) =
		\ope{A}_{t_0}^{\mathrm{H}}(t_0) = \ope{A}(t_0)
	\end{equation}
but now $\ope{A}(t_0)$ is generally different from $\ope{A}(t)$.
In this way we have proved that an
\emph{%
observable is an integral of motion iff in the Heisenberg picture it
coincides with its initial value in the Schr\"odinger picture.%
}

	As in the explicitly time-independent case considered above, now one
can easily prove that if some state vector is an eigenvector for $\ope{A}$
with an eigenvalue $a$, than $\ope{A}$ is an integral of motion iff $a$ is
time-independent, i.e.\ $a=\const$.%
\footnote{%
In fact, in this case we have $\ope{A}(t)\psi(t)=a(t)\psi(t)$ for $\psi(t)$
satisfying $\ih\frac{d\psi(t)}{d t}=\Ham(t)\psi(t)$. The integrability
condition for this system of equations (with respect to $\psi(t)$) is
\(
\ih\frac{\partial\ope{A}(t)}{\partial t} +[\ope{A}(t),\Ham(t)]_{\_} =
\ih\frac{d a(t)}{d t} \id_\Hil
\)
from where the above result follows.%
}

	Now we are going to consider the problem of integrals of motion from
the fibre bundle view-point.

	A dynamical variable is called an
\emph{integral of motion} if the corresponding to it bundle morphism
(see Sect.~\ref{VI}) has time-independent meanvalues, viz.\
	\begin{equation}	\label{8.11}
\langle \mor{A}_\gamma(t)\rangle_{\Psi_\gamma}^{t} =
\langle \mor{A}_\gamma(t_0)\rangle_{\Psi_\gamma}^{t_0}
	\end{equation}
which, due to~(\ref{6.1''}) is equivalent (and equal) to~(\ref{8.5}).
From~(\ref{8.6}), (\ref{4.4}), (\ref{6.1'}), (\ref{4.11}), (\ref{4.14}),
and~(\ref{7.12}), we see that~(\ref{8.11}) us equivalent to
	\begin{equation}	\label{8.12}
\mor{A}_\gamma(t) =
\mor{U}_\gamma(t,t_0)\circ \mor{A}_\gamma(t_0) \circ \mor{U}_\gamma(t_0,t) =
\mor{A}_{\gamma,t_0}^{\mathrm{H}}(t).
	\end{equation}
A feature of the Hilbert bundle description is that in it, for the difference
of the Hilbert space one, we cannot directly differentiate with respect to $t$
maps like
\(
\mor{A}_\gamma(t)\colon \bHil_{\gamma(t)} \to \bHil_{\gamma(t)}
\)
and
\(
\mor{U}_\gamma(t,t_0)\colon \bHil_{\gamma(t_0)} \to \bHil_{\gamma(t)} .
\)
	So, to obtain the differential form of~(\ref{8.12})
(or~(\ref{8.11})) we differentiate with respect to $t$ the matrix form
of~(\ref{8.12}) in a given field of bases (see Sect.~\ref{V}).
Thus, using~(\ref{5.6}), we find
	\begin{equation}	\label{8.13}
\ih \frac{\partial\mmor{A}_\gamma(t)}{\partial  t} +
\left[
{\mmor{A}_\gamma(t)} , {\mbHam_{\gamma}(t)}
\right]_{\_}
 = 0.
	\end{equation}
Because of~(\ref{5.10}) and~(\ref{6.5}) this equation is the local
matrix form of the invariant equation
	\begin{equation}	\label{8.14}
\widetilde{\mor{D}}_{t}^{\gamma}\left(\mor{A}_\gamma\right)=0.
	\end{equation}

	Consequently a
\emph{%
dynamical variable is an integral of motion iff the corresponding to it
bundle
morphism has a vanishing derivative along any observer's world line with
respect to the derivation along paths (of the fibre morphisms) associated
with the bundle evolution transport.%
}

	If in some basis
\( \mmor{A}_\gamma(t)=\const=\mmor{A}_\gamma(t_0) \),
then, with the help of~(\ref{8.8}), we get
\(
[
\mmor{A}_\gamma(t) , \mbHam_{\gamma}(t)
]_{\_}
= 0,
\)
i.e.\ the matrix of $\mor{A}_\gamma$
and the matrix-bundle Hamiltonian commute.
It is important to note that from here does not follow the commutativity of
the morphism along paths, representing an observable by~(\ref{6.0}),
and the bundle Hamiltonian~(\ref{6.2'}) because the matrix of the latter
is connected with the matrix-bundle Hamiltonian through~(\ref{6.2''}).

	If the state vector $\Psi_\gamma(t)$ is an eigenvector for
$\mor{A}_\gamma(t)$, i.e.\
\(
\mor{A}_\gamma(t)\Psi_\gamma(t) = a(t) \Psi_\gamma(t),\
a(t)\in\mathbb{R}
\),
then $\langle \mor{A}_\gamma(t)\rangle_{\Psi_\gamma}^{t} = a(t)$.
Hence from~(\ref{8.6}) follows that $\mor{A}_\gamma$  is an integral of
motion iff $a(t)=\const=a(t_0)$.

	Rewriting equation~\eref{8.13} in the form of Lax pair
equation~\cite{Lax}
	\begin{equation}	\label{Lax-equation}
\frac{\partial}{\partial t} \mmor{A}_\gamma(t) =
- \iih [ \mmor{A}_\gamma(t) ,\mHam_\gamma(t) ]_- =
[ \mmor{A}_\gamma(t) , \boldsymbol{\Gamma}_\gamma(t) ]_- ,
	\end{equation}
where~\eref{5.10} was taken into account, we see that
\emph{%
$\dyn{A}$ is an
integral of motion iff in some (and hence in any) field of bases the matrices
$\mmor{A}_\gamma(t)$ and $\boldsymbol{\Gamma}_\gamma(t)$ form a Lax pair.%
}

	It is known~\cite[sect.~2]{Rosquist&Goliath} that the Lax pair
equation~\eref{Lax-equation} is invariant under  transformations of a form
	\begin{equation}	\label{Lax-invariance}
\mmor{A}_\gamma(t) \mapsto \mmor{W}\mmor{A}_\gamma(t)\mmor{W}^{-1}, \qquad
\boldsymbol{\Gamma}_\gamma(t) \mapsto
	\mmor{W}\boldsymbol{\Gamma}_\gamma(t)\mmor{W}^{-1} -
	\frac{\partial\mmor{W}}{\partial t}\mmor{W}^{-1}
	\end{equation}
where $\mmor{W}$ is a nondegenerate matrix, possibly depending on $\gamma$
and $t$ in our case. Hence $\mmor{A}_\gamma(t)$ transforms as a tensor while
$\boldsymbol{\Gamma}_\gamma(t)$ transforms as a connection. These
observations fully agree with our results of Sect.~\ref{V}, expressed by
equations~\eref{5.02b} and~\eref{5.11} with
$(\boldsymbol{\Omega}^\top(t;\gamma))^{-1}=\mmor{W}$,
and give independent arguments for treating (up to a constant) the
matrix-bundle Hamiltonian as a gauge (connection) field.

	Let us consider briefly the bundle description of integrals
of motion in the Heisenberg picture. In it~(\ref{8.11}) transforms to
	\begin{equation}	\label{8.15}
\langle \mor{A}_{\gamma,t}^{\mathrm{H}}(t_0) \rangle_{\Psi_\gamma}^{t_0} =
\langle \mor{A}_{\gamma,t_0}^{\mathrm{H}}(t_0) \rangle_{\Psi_\gamma}^{t_0}
	\end{equation}
which is equivalent to (cf.~(\ref{8.10}))
	\begin{equation}	\label{8.16}
\mor{A}_{\gamma,t}^{\mathrm{H}}(t_0) = \mor{A}_{\gamma,t_0}^{\mathrm{H}}(t_0)
\ \left( = \mor{A}_\gamma(t_0) \right) .
	\end{equation}
So, due to~(\ref{7.10}), the observable $\mor{A}$ is an integral of
motion iff (cf.~(\ref{8.9}))
	\begin{equation}	\label{8.17}
0 = \ih \frac{\partial \mor{A}_{\gamma,t}^{\mathrm{H}}(t_0)}{\partial t} =
\left[
\mor{A}_{\gamma,t}^{\mathrm{H}}(t_0) , \bHam_{\gamma,t}^{\mathrm{H}}(t_0)
\right]_{\_}
+
\ih
\left(
\frac{\partial \ope{A}}{\partial t}
\right)_{\! \gamma,t}^{\!\mathrm{H}}(t_0).
	\end{equation}

	This equation, due to~\eqref{7.13b7}, is equivalent to
(cf.~\eqref{8.14})
	\begin{equation}	\label{8.17a}
^\circ{\!}\mor{D}_{t}^{\gamma}(\mor{A}_\gamma) = 0.
	\end{equation}

	Therefore a
\emph{%
dynamical variable is an integral of motion iff the corresponding to it
morphism along paths of $\HilB$ has a vanishing derivative along any observer's
trajectory with respect to the derivation along paths in $\morf{\HilB}$
induces by the evolution transport (along paths).%
}

	Ending this section, we note that the description of the integrals of
motion in the Hilbert space $\Hil$ and in the Hilbert fibre bundle $\HilB$
are fully equivalent because of~(\ref{6.0}) and~(\ref{4.3}).

\section{Conclusion}
\label{conclusion-III}
\setcounter{equation}{0}

	As we have seen, the different pictures of motion (Schr\"odinger,
Heisenberg, etc.) of quantum mechanics have their natural analogues in the
fibre bundle approach to it. Any one of them simplifies one or the other aspect
of the theory and is suitable for consideration of corresponding concrete
problems. The integrals of motion, investigated here from a fibre bundle
view-point, are a typical example of this kind in which the Heisenberg
picture of motion is the most suitable one. We have derived invariant (fibre
bundle) necessary and sufficient conditions for a dynamical variable to be an
integral of motion.

	Further in this series we intend to consider problems connected with
fibre bundle description of mixed states, evolution transport's curvature,
interpretation of the Hilbert bundle description of quantum mechanics and its
possible developments.


\addcontentsline{toc}{section}{References}

\bibliography{bozhopub,bozhoref}
\bibliographystyle{unsrt}

\addcontentsline{toc}{subsubsection}{\vspace{1ex}This article ends at page}

\end{document}